\documentclass[12pt]{article}

\begin{document}
%\selectlanguage{english}

\baselineskip 0.76cm
\topmargin -0.4in
\oddsidemargin -0.1in

\let\ni=\noindent

\renewcommand{\thefootnote}{\fnsymbol{footnote}}

\newcommand{\SM}{Standard Model }

\pagestyle {plain}

\setcounter{page}{1}

%\pagestyle{empty}

%\addtocounter{equation}{+1}

~~~~~
\pagestyle{empty}

\begin{flushright}
IFT-02/08 
\end{flushright}

\vspace{0.3cm}

{\large\centerline{\bf Clifford algebra implying three fermion generations 
revisited{\footnote {Work supported in part by the Polish State Committee 
for Scientific Research (KBN), grant 5 P03B 119 20 (2001--2002).}}}}

\vspace{0.5cm}

{\centerline {\sc Wojciech Kr\'{o}likowski}}

\vspace{0.23cm}

{\centerline {\it Institute of Theoretical Physics, Warsaw University}}

{\centerline {\it Ho\.{z}a 69,~~PL--00--681 Warszawa, ~Poland}}

\vspace{0.3cm}

{\centerline{\bf Abstract}}

\vspace{0.2cm}

The author's idea of {\it algebraic compositeness} of fundamental particles, 
allowing to understand the existence in Nature of three fermion generations, 
is revisited. It is based on two postulates.~({\it i}) For all fundamental 
particles of matter the Dirac square-root procedure 
$\sqrt{p^2}\! \rightarrow \!\Gamma^{(N)}\cdot p$ works, leading to a 
sequence $N = 1,2,3,\ldots$ of Dirac-type equations, where four Dirac-type 
matrices $\Gamma^{(N)}_\mu $ are embedded into a Clifford algebra {\it via} 
a Jacobi definition introducing four "centre-of-mass" and $(N-1) \times $four
"relative" Dirac-type matrices. These define one "centre-of-mass" and $N-1$ 
"relative" Dirac bi\-spi\-nor indices.~({\it ii}) The "centre-of-mass" Dirac 
bispi\-nor index is coupled to the \SM gauge fields, while $N-1$ "relative" 
Dirac bispinor indices are all free undistinguishable physical objects 
obeying Fermi statistics with the Pauli principle which requires the full 
antisymmetry with respect to "relative" indices. This allows only for 
{\it three} $N = 1,3,5$ in the case of $N$ odd, implying the existence of 
{\it three and only three} generations of fundamental fermions.

\vspace{0.3cm}

\ni PACS numbers: 12.50.Ch , 12.90+b , 12.10.Dm 

\vspace{0.6cm}

\ni March 2002

\vfill\eject

~~~~~

\pagestyle {plain}

\setcounter{page}{1}

\vspace{0.2cm}

One of the most important theoretical achievements in the history of physics 
was Dirac's algebraic discovery of the particle's spin 1/2, inherently 
connected with the linearization of relativistic wave equation through his 
fameous square-root procedure $\sqrt{p^2} \rightarrow \gamma\cdot p$ 
\cite{DIR}. 
As a result, some physical particles, called later on spin-1/2 fermions, got 
--- in addition to their spatial coordinates $\vec{r}$ --- new algebraic 
degrees of freedom, described with the use of Dirac bispinor index 
$\alpha = 1,2,3,4$. This was acted on by the $4\times 4$ Dirac matrices, 
in particular, by the spin-1/2 matrix $\frac{1}{2}\vec{\sigma}$ which 
supplemented the orbital angular momentum operator $\vec{r}\times \vec{p}$ 
to the operator of particle's total angular momentum. Thus, through an act 
of abstraction from the particle's spatial properties, its spin 1/2 was 
recognized as an algebraic analogue of its spatial, orbital angular momentum,
 satisfying the same rotation-group commutation relations.

The starting point of the author's idea of {\it algebraic compositeness} 
\cite{KR} was a proposal of a new act of abstraction from the particle's 
spatial properties, where a wave function 
$\psi^{(N)}_{\alpha_1 \alpha_2 \ldots \alpha_N}(\vec{r})$ of several Dirac 
bispinor indices $\alpha_1, \alpha_2, \ldots, \alpha_N$ (as representing $N$ 
physical objects correlated with one material point of spatial coordinates 
$\vec{r}$) was introduced in an analogy with the wave function 
$\psi^{(N)}(\vec{r}_1 ,\vec{r}_2 ,\ldots, \vec{r}_N)$ of several spatial 
coordinates $ \vec{r}_1, \vec{r}_2,\ldots,\vec{r}_N$ (as representing $N$ 
physical objects being, this time, material points).

In order to construct these several Dirac bispinor indices the observation 
was made \cite{KR} that, generically, the Dirac algebra

% (1)
\begin{equation}
\left\{ \Gamma^{(N)}_\mu\,,\,\Gamma^{(N)}_\nu \right\} = 2 g_{\mu \nu}
\end{equation}

\ni can be realized through Dirac-type matrices of the form 

%(2)
\begin{equation}
\Gamma^{(N)}_\mu \equiv \frac{1}{\sqrt{N}} \sum^N_{i=1}  \gamma^{(N)}_{i \mu} 
\end{equation}

\ni built up linearly from $N$ elements of the Clifford algebra

%(3)
\begin{equation}
\left\{ \gamma^{(N)}_{i \mu}\,,\,\gamma^{(N)}_{j \nu} \right\} 
= 2 \delta_{i j} g_{\mu \nu}\,, 
\end{equation}

\ni  where $N = 1,2,3,\ldots\;,\; i,j = 1,2,\ldots,N$ and 
$\mu,\nu = 0,1,2,3$. Then, the Dirac square-root procedure 
$\sqrt{p^2} \rightarrow \Gamma^{(N)}\cdot p$ leads in the interaction-free 
case to the sequence $N = 1,2,3,\ldots $ of Dirac-type equations

%(4)
\begin{equation}
\left( \Gamma^{(N)}\cdot p -  M^{(N)}\right) \psi^{(N)}(x) = 0
\end{equation}

\ni with $\psi^{(N)}(x) \equiv 
\left(\psi^{(N)}_{\alpha_1 \alpha_2 \ldots \alpha_N}(x)\right)$, where the 
meaning of Dirac bispinor indices is explained as in Eq. (12) later on. 
Here, each $\alpha_i = 1,2,3,4 \;\;(i =1,2,\ldots,N)$. The mass $ M^{(N)}$ 
is independent of $\Gamma^{(N)}_\mu$. In general, the mass $ M^{(N)}$ should 
be replaced by a mass matrix of elements $ M^{(N,N')}$ which would couple 
$\psi^{(N)}(x)$ with all appropriate $\psi^{(N')}(x)$, and it might be 
natural to assume for $N \neq N'$ that $\gamma^{(N)}_{i \mu}$ and 
$\gamma^{(N')}_{j \nu}$ commute, and so do $\Gamma^{(N)}_\mu $ and 
$\Gamma^{(N')}_\nu $.

For $ N = 1$, Eq. (4) is evidently the usual Dirac equation and for $ N = 2$ 
it is known as the Dirac form \cite{BA} of  K\"{a}hler equation \cite{KA}, 
while for $ N \geq 3$ Eqs. (4) give us {\it new} Dirac-type equations 
\cite{KR}. They describe some spin-halfinteger or spin-integer particles 
for $N$ odd or $N$ even, respectively. 

The Dirac-type matrices $\Gamma^{(N)}_\mu $ for any $N$ can be embedded 
into the new Clifford algebra

%(5)
\begin{equation}
\left\{ \Gamma^{(N)}_{i \mu}\, ,  \,\Gamma^{(N)}_{j \nu} \right\} 
= 2\delta_{i j} g_{\mu \nu}\;,
\end{equation}

\ni isomorphic with the Clifford algebra (3) of $\gamma^{(N)}_{i \mu}$, if 
$\Gamma^{(N)}_{i \mu}$ are defined by the properly normalized Jacobi linear 
combinations of $\gamma^{(N)}_{i \mu}$:

%(6)
\begin{eqnarray}
\Gamma^{(N)}_{1 \mu} & \equiv & \Gamma^{(N)}_\mu \equiv 
\frac{1}{\sqrt{N}} 
\left(\gamma^{(N)}_{1 \mu} + \ldots + \gamma^{(N)}_{N \mu} \right) \,, 
\nonumber \\ 
\Gamma^{(N)}_{i \mu} & \equiv & \frac{1} {\sqrt{i(i - 1)}} 
\left[ \gamma^{(N)}_{1 \mu} + \ldots + 
\gamma^{(N)}_{i\!-\!1\, \mu} - (i - 1) \gamma^{(N)}_{i \mu} \right] 
\end{eqnarray}

\ni for $ i = 1$ and $ i = 2,\ldots, N$, respectively. So, 
$\Gamma^{(N)}_{1 \mu}$ and 
$ \Gamma^{(N)}_{2 \mu},\ldots,\Gamma^{(N)}_{N \mu}$, respectively, present  
the  "centre-of-mass"\, and "relative"\, Dirac-type matrices. Note that the 
Dirac-type equation (4) for any $N$ does not involve the "relative"\, 
Dirac-type matrices 
$\Gamma^{(N)}_{2 \mu} \, , \, \ldots, \Gamma^{(N)}_{N \mu}$, including 
solely the "centre-of-mass"\, Dirac-type matrix 
$\Gamma^{(N)}_{1 \mu} \equiv \Gamma^{(N)}_\mu $. Since 
$\Gamma^{(N)}_{i \mu} = \sum^N_{j=1} O_{i j} \gamma^{(N)}_{j \mu}$, where 
$ O = \left( O_{i j} \right)$ is an orthogonal $N\times N$ matrix 
($O^T = O^{-1}$), we obtain for the total spin tensor the equality

%(7)
\begin{equation}
\sum^N_{i=1}  \sigma^{(N)}_{i \mu \nu} = \sum^N_{i=1}  
\Sigma^{(N)}_{i \mu \nu}  \,,
\end{equation}

\ni where

%(8)
\begin{equation} 
\sigma^{(N)}_{j \mu \nu} \equiv \frac{i}{2} 
\left[ \gamma^{(N)}_{j \mu} \, , \, 
\gamma^{(N)}_{j \nu} \right] \;\; ,\;\; 
\Sigma^{(N)}_{j \mu \nu} \equiv \frac{i}{2} 
\left[ \Gamma^{(N)}_{j \mu} \, , \, \Gamma^{(N)}_{j \nu} \right] \,. 
\end{equation}

\ni The total spin tensor (7) is the generator of Lorentz 
transformations for $\psi^{(N)}(x)$.

In place of the chiral representations for individual 
$\gamma^{(N)}_ j = \left(\gamma^{(N)}_{j \mu} \right)$, where   

\vspace{-0.1cm}

%(9)
\begin{equation}
\gamma^{(N)}_{j 5} \equiv i \gamma^{(N)}_{j 0} \gamma^{(N)}_{j 1} 
\gamma^{(N)}_{j 2} \gamma^{(N)}_{j 3} \;\; , \;\; \sigma^{(N)}_{j 3} 
\equiv \sigma^{(N)}_{j 12}
\end{equation}

\ni are diagonal, it is convenient to use for any $N$ the chiral 
representations of Jacobi 
$\Gamma^{(N)}_j = \left(\Gamma^{(N)}_{j \mu} \right)$, where now

%(10)
\begin{equation}
\Gamma^{(N)}_{j 5} \equiv i\Gamma^{(N)}_{j 0} \Gamma^{(N)}_{j 1} 
\Gamma^{(N)}_{j 2} \Gamma^{(N)}_{j 3} \;\; ,\;\; \Sigma^{(N)}_{j 3} 
\equiv  \Sigma^{(N)}_{j 12}
\end{equation}

%\vspace{0.15cm}

\ni are diagonal (all matrices (9) and similarly (10) commute 
simultaneously, both with equal and different $j$). 

When using the Jacobi chiral representations, the "centre-of-mass"\, 
Dirac-type matrices  $\Gamma^{(N)}_{1 \mu} \equiv \Gamma^{(N)}_\mu$ 
and $ \Gamma^{(N)}_{1 5} \equiv \Gamma^{(N)}_5 \equiv i\Gamma^{(N)}_0 
\Gamma^{(N)}_1 \Gamma^{(N)}_2 \Gamma^{(N)}_3$ can be taken in the 
reduced forms

%(11)
\begin{equation}
\Gamma^{(N)}_\mu =  \gamma_\mu \otimes 
\underbrace{ {\bf 1}\otimes \cdots \otimes {\bf 1}}_{ N-1 \;{\rm times}} 
\;\; , \;\; \Gamma^{(N)}_5 = \gamma_5  \otimes 
\underbrace{ {\bf 1}\otimes \cdots \otimes {\bf 1}}_{ N-1 \;{\rm times}} \; , 
\end{equation}

\ni where $\gamma_\mu$, $ \gamma_5 \equiv i \gamma_0 \gamma_1 \gamma_2 
\gamma_3 $ and {\bf 1} are the usual $4\times 4$ Dirac matrices. 

Then, the Dirac-type equation (4) for any $N$ can be rewritten in the 
reduced form

%(12)
\begin{equation}
\left( \gamma \cdot p  - M^{(N)}\right)_{\alpha_1\beta_1} 
\psi^{(N)}_{\beta_1 \alpha_2 \ldots \alpha_N}(x) = 0\;,
\end{equation}

\ni where $\alpha_1$ and $\alpha_2 \,,\, \ldots\,,\, \alpha_N$ are 
the "centre-of-mass"\, \,and "relative"\, \,Dirac bispinor indices, 
respectively ($\alpha_i =1,2,3,4$ for any $i = 1,2,\ldots,N$). Note 
that in the Dirac-type equation (12) for any $N>1$ there appear the 
"relative"\, Dirac indices $\alpha_2 \,,\, \ldots\,,\, \alpha_N$ which are 
free from any coupling, but still are subjects of Lorentz transformations. 

The Standard Model gauge interactions can be introduced to the Dirac-type 
equations (12) by means of the minimal substitution 
$p \rightarrow p - g A(x)$, where $p$ plays the role of 
the "centre-of-mass" \,four-momentum, and so, $x$ --- the 
"centre-of-mass" \,four-position. Then,

%(13)
\begin{equation}
\left\{ \gamma \cdot \left[p - g A(x)\right] - 
M^{(N)}\right\}_{\alpha_1\beta_1} 
\psi^{(N)}_{\beta_1 \alpha_2 \ldots \alpha_N}(x) = 0\;,
\end{equation}

\ni where $g \gamma \cdot A(x)$ symbolizes the Standard Model gauge 
coupling that involves within $A(x)$ the familiar weak-isospin and color 
matrices, the weak-hypercharge dependence as well as the usual Dirac chiral 
matrix $\gamma_5 $. The last arises from the "centre-of-mass" \, Dirac-type 
chiral matrix $\Gamma^{(N)}_5 $, when a generic 
$g \Gamma^{(N)}\! \cdot \!A(x)$ is reduced to $g \gamma\!  \cdot \! A(x)$  
in Eqs. (13) [see Eq. (11)]. Note that then 
$A_\mu(x) \equiv A_\mu(x,\gamma_5) \equiv A_\mu(x,0) + A'_\mu(x,0)\gamma_5$ 
depends linearly on $\gamma_5$.

In Eqs. (13) the Standard Model gauge fields interact only with the 
"centre-of-mass" \,index $\alpha_1$ that, therefore, is distinguished from 
the physically unobserved "relative" \,indices $\alpha_2, \ldots, \alpha_N$. 
This was the reason, why some time ago we conjectured that the 
"relative" \,Dirac bispinor  indices $\alpha_2, \ldots, \alpha_N$ are all 
undistinguishable physical objects obeying Fermi statistics along with the 
Pauli principle requiring the full antisymmetry of wave function 
$\psi^{(N)}_{\alpha_1 \alpha_2 \ldots \alpha_N}(x)$ with respect to 
$\alpha_2, \ldots, \alpha_N$ \cite{KR}. Hence, due to this "intrinsic Pauli 
principle", only five values of $N$ satisfying the condition $N-1\leq 4$ 
are allowed, namely $N = 1,3,5$ for $N$ odd and $N = 2,4$ for $N$ even. 
Then, from the postulate of relativity and the probabilistic interpretation 
of $\psi^{(N)}(x) \equiv 
\left(\psi^{(N)}_{\alpha_1 \alpha_2 \ldots \alpha_N}(x)\right) $ we were 
able to infer that these $N$ odd and $N$ even correspond to states with 
total spin 1/2 and total spin 0, respectively \cite{KR}.

Thus, the Dirac-type equation (13), jointly with the "intrinsic Pauli 
principle", if considered on a fundamental level, justifies the existence 
in Nature of {\it three and only three} generations of spin-1/2 fundamental 
fermions coupled to the \SM gauge bosons  (they are identified with leptons 
and quarks). In addition, there should exist {\it two and only two} 
generations of spin-0 fundamental bosons also coupled to the \SM gauge 
bosons (they are not identified yet). Note that one cannot hope here for 
a construction of the full supersymmetry. At most, there might appear a 
partial supersymmetry: two to two, broken by the absence of one boson 
generation (the question is of which).

The wave functions or fields of spin-1/2 fundamental fermions (leptons 
and quarks) of three generations $N = 1,3,5$ can be presented in terms 
of $\psi^{(N)}_{\alpha_1 \alpha_2 \ldots \alpha_N}(x)$ as follows:

%(14)
\begin{eqnarray} 
\psi^{(f_1)}_{\alpha_1}(x) & = & \psi^{(1)}_{\alpha_1}(x) \;, \nonumber \\
\psi^{(f_3)}_{\alpha_1}(x) & = & 
\frac{1}{4}\left(C^{-1} \gamma_5 \right)_ {\alpha_2 \alpha_3} 
\psi^{(3)}_{\alpha_1 \alpha_2 \alpha_3}(x) = \psi^{(3)}_{\alpha_1 1 2}(x) = 
\psi^{(3)}_{\alpha_1 3 4}(x) \;,\nonumber \\
\psi^{(f_5)}_{\alpha_1}(x) & = & \frac{1}{24}
\varepsilon_{\alpha_2 \alpha_3 \alpha_4 \alpha_5} 
\psi^{(5)}_{\alpha_1 \alpha_2 \alpha_3 \alpha_4 \alpha_5}(x) = 
\psi^{(5)}_{\alpha_1 1 2 3 4}(x) \;,
\end{eqnarray}  

\ni where $ \psi^{(N)}_{\alpha_1 \alpha_2 \ldots \alpha_N}(x) $ carries 
also the \SM (composite) label, suppressed in our notation, and $C$ denotes 
the usual $4\times 4$ charge-conjugation matrix. Here, writing explicitly, 
$f_1 = \nu_e\,,\,e^-\,,\, u\,,\,d \;,\; f_3 = \nu_\mu\,,\, 
\mu^-\,,\, c\,,\,s $ and $f_5 = \nu_\tau\,,\, \tau^-\,,\,t \,,\,b $, 
thus each $f_N$ corresponds to the same suppressed \SM (composite) label. 
We can see that, due to the full antisymmetry in $\alpha_i $ indices for 
$i \geq 2$, the wave functions or fields $N =1,3$ and 5 appear (up to the 
sign) with the multiplicities 1, 4 and 24,  respectively. Thus, for them, 
there is defined the weighting matrix

%(15)
\begin{equation} 
\rho^{1/2} = \frac{1}{\sqrt{29}}  
\left( \begin{array}{ccc}  1  & 0 & 0 \\ 
0 & \sqrt4 & 0  \\ 0 & 0 & \sqrt{24} \end{array} \right) \;, 
\end{equation}

\ni where Tr $\rho = 1$.

For each bispinor wave function or field 
$\psi^{(f_N)}_{\alpha_1}(x)\;\;(N = 1,3,5)$ defined in Eqs. (14), the 
Dirac-type equation (13) can be reduced to the usual Dirac equation 

%(16)
\begin{equation}
\left\{ \gamma \cdot \left[ p-g A(x)\right] - M^{(N)}\right\}_{\alpha_1 
\beta_1} \psi^{(f_N)}_{\beta_1}(x) = 0 \;. 
\end{equation}

\ni This gives in turn the relativistic covariant conserved current of 
the usual Dirac form

%(17)
\begin{equation}
j^{(f_N)}_{\mu {\rm D}}(x) \equiv \psi^{(f_N) *}_{\alpha_1}(x) 
\left(\gamma_0 \gamma_\mu\right)_{\alpha_1 \beta_1} 
\psi^{(f_N)}_{\beta_1}(x)  \,.
\end{equation}

\ni In fact, $\partial^\mu j^{(f_N)}_{\mu {\rm D}}(x) = 0$ since 
$ A^\dagger_\mu (x) =  A_\mu (x)$ where $\gamma_5^\dagger = \gamma_5 $. 

Concluding the first part of this note, we would like to point out that 
our algebraic construction of {\it three and only three} generations of 
leptons and quarks may be interpreted {\it either} as ingenuously algebraic 
(much like the famous Dirac's algebraic discovery of spin 1/2), {\it or} as 
a summit of an iceberg of really composite states of $N$ spatial partons 
with spin 1/2 whose Dirac bispinor indices manifest themselves as our Dirac 
bispinor  indices $\alpha_1, \alpha_2, \ldots, \alpha_N$ ($N = 1,3,5$) which 
thus may be called "algebraic partons", as being algebraic building blocks 
for leptons and quarks. Among all $N$ "algebraic partons" \,in any 
generation $N$ of leptons and quarks, there are one "centre-of-mass 
algebraic parton" \,$ (\alpha_1)$ and $ N-1$ "relative algebraic partons" 
\,$(\alpha_2, \ldots, \alpha_N)$, the latter undistinguishable from each 
other and so, obeying our "intrinsic Pauli principle".  

Now, we pass to some more formal discussion. It is not difficult to see 
that both for $N$ odd and $N$ even the Dirac-type equation (13) implies 
the local conservation of the following relativistic covariant structure:

%18
\begin{equation}
j^{(N)}_{\mu \alpha_2 \ldots \alpha_N, \beta_2 \ldots \beta_N}(x) \equiv 
\psi^{(N)*}_{\alpha_1\alpha_2 \ldots \alpha_N}(x) 
\left(\gamma_0 \gamma_\mu \right)_{\alpha_1  \beta_1} 
\psi^{(N)}_{\beta_1 \beta_2 \ldots \beta_N}(x) \,.  
\end{equation}
%w238

\ni In fact, $ \partial^\mu j_{\mu \alpha_2 \ldots \alpha_N, 
\beta_2 \ldots \beta_N}(x) = 0$ because $A^\dagger_\mu(x) = A_\mu(x)$. 
The local conservation of the currents (17) for $N = 1,3,5$ follows 
immediately from Eq. (18), since

%19
\begin{eqnarray}
j^{(f_1)}_{\mu {\rm D}}(x) & = & j^{(1)}_{\mu}(x)\,, \nonumber \\
j^{(f_3)}_{\mu {\rm D}}(x) & = & \frac{1}{4} 
\left( C^{-1} \gamma_{5} \right)^*_{\alpha_2 \alpha_3} 
j^{(3) }_{\mu \alpha_2 \alpha_3, \beta_2 \beta_3}(x)
\frac{1}{4}\left(C^{-1} \gamma_{5} \right)_{\beta_2 \beta_3} \,, \nonumber \\
j^{(f_5)}_{\mu {\rm D}}(x) & = & \frac{1}{24} 
\varepsilon_{\alpha_2 \alpha_3 \alpha_4 \alpha_5} 
j^{(5) }_{\mu \alpha_2 \alpha_3 \alpha_4 \alpha_5, 
\beta_2 \beta_3 \beta_4 \beta_5}(x)\frac{1}{24} 
\varepsilon_{\beta_2 \beta_3 \beta_4 \beta_5} \,. 
\end{eqnarray}

In general, the relativistic covariant Dirac-type currents both 
for $N$ odd and $N$ even must have the form

%20
\begin{equation}
j^{(N)}_{\mu {\rm D}}(x) \equiv \psi^{(N) *}_{\alpha_1 
\alpha_2 \ldots \alpha_N}(x) \xi^{(N)} \left( \Gamma^{(N)}_{10} 
\Gamma^{(N)}_{20} \ldots\Gamma^{(N)}_{N0} 
\Gamma^{(N)}_{1\mu} \right)_{\alpha_1 \alpha_2 
\ldots \alpha_N, \beta_1 \beta_2 \ldots \beta_N} 
\psi^{(N)}_{\beta_1 \beta_2 \ldots \beta_N}(x) \,,
\end{equation}

\ni where $\Gamma^{(N)}_{i\mu}$ are the Dirac-type matrices in their 
Jacobi version, introduced in Eqs. (6), while $\xi^{(N)}$ is a phase 
factor making Hermitian the $N\times N$ bispinor matrix appearing in 
this current. For $N$ even this definition is trivial, as then the 
Dirac-type current vanishes. In the case of $N$ odd, we are going 
to show that

%21
\begin{equation}
j^{(N)}_{\mu {\rm D}}(x) =j^{(N)}_{\mu \alpha_2 \ldots 
\alpha_N , \beta_2 \ldots \beta_N}(x) 
(\gamma_{0})_{\alpha_2 \beta_2}\ldots (\gamma_{0})_{\alpha_N \beta_N} \,. 
\end{equation}

\ni Thus, $\partial^\mu j^{(N)}_{\mu {\rm D}}(x) = 0$ for $N$ odd.

To prove Eq. (21), we observe that the Dirac-type matrices 
$\Gamma^{(N)}_{i\mu}\;\;(i = 1,2, \ldots,N)$, satisfying the 
anticommutation relations of Clifford algebra (5), can be represented 
in terms of the usual $4\times 4$ Dirac matrices as follows:

%22
\begin{eqnarray}
\Gamma^{(N)}_{1\mu} & = & \gamma_\mu \otimes \,\underbrace{{\bf 1}\, 
\otimes \,{\bf 1}\, \otimes \,{\bf 1}\, \otimes \!\ldots 
\otimes {\bf 1} \otimes {\bf 1}}_{N-1\;{\rm times}} \,, \nonumber \\
\Gamma^{(N)}_{2\mu} & = & \gamma_5 \otimes \!i\gamma_\mu \! \gamma_5\! 
\otimes {\bf 1} \otimes {\bf 1} \otimes \!\ldots \otimes {\bf 1} 
\otimes {\bf 1} \,, \nonumber \\
\Gamma^{(N)}_{3\mu} & = & \gamma_5 \otimes \,\gamma_5 \otimes 
\,\gamma_\mu \otimes {\bf 1} \otimes \ldots \otimes {\bf 1} 
\otimes {\bf 1} \,, \nonumber \\
.\;\;.\;\;.\!\! & .\;\;. 
& \!\!.\;\;.\;\;.\;\;.\;\;.\;\;.\;\;.\;\;.\;\;.
\;\;.\;\;.\;\;.\;\;.\;\;.\;\;.\;\;.\;\;. \nonumber \\
\Gamma^{(N)}_{N\mu} & = & \underbrace{\gamma_5 \otimes \gamma_5 
\otimes \gamma_5 \otimes \gamma_5 \otimes \ldots \otimes 
\gamma_5}_{N-1\;{\rm times}} \otimes \left\{ \begin{array}{ll} 
\gamma_\mu & {\rm for}\;N\;{\rm odd} \\ i \gamma_\mu \gamma_5 & 
{\rm for}\;N\;{\rm even} \end{array} \right. \,,
\end{eqnarray}

\ni what is an extension of the representation (11) for 
$\Gamma^{(N)}_{1\mu} \equiv \Gamma^{(N)}_\mu $ leading to the form (13) 
of Dirac-type equation for any $N$.  Forming their product for $\mu = 0 $,

%23
\begin{equation}
\Gamma^{(N)}_{10} \Gamma^{(N)}_{20} \ldots \Gamma^{(N)}_{N0} = 
\left\{ \begin{array}{cl} \;i^{\frac{N-1}{2}}\;\;\;\gamma_0\;\, 
\otimes \;\,\gamma_0\;\, \otimes \ldots \otimes \;\,\gamma_0 & 
{\rm for}\;N\;{\rm odd} \\
(- i)^{\frac{N}{2}} i\gamma_0 \gamma_5\! \otimes \!i\gamma_0 
\gamma_5\! \otimes \!\ldots \otimes \! i\gamma_0 \gamma_5\! & 
{\rm for}\;N\;{\rm even} \end{array} \right.\,, 
\end{equation}

\ni and multiplying from the right by $\Gamma^{(N)}_{1 \mu}$, we obtain

%24
\begin{equation}
\Gamma^{(N)}_{10} \Gamma^{(N)}_{20} \ldots \Gamma^{(N)}_{N0}
\Gamma^{(N)}_{1\mu} = \left\{ \begin{array}{cl} \;\;i^{\frac{N-1}{2}}\;\, 
\gamma_0\, \gamma_\mu\;\,\otimes \;\,\gamma_0\;\, \otimes \ldots 
\otimes \;\, \gamma_0 & {\rm for}\;N\;{\rm odd} \\
(- i)^{\frac{N-2}{2}} \gamma_0\! \gamma_5\! \gamma_\mu\! \otimes 
\!i\gamma_0 \gamma_5\! \otimes \! \ldots \otimes\! i\gamma_0 
\gamma_5\! & {\rm for}\,N\,{\rm even} \end{array} \right.\,. 
\end{equation}

\ni Hence, we can define the phase factors in Eq. (20) as follows:

%25
\begin{equation}
\xi^{(N)} =  \left\{ \begin{array}{ll} (-i)^{\frac{N-1}{2}} & 
{\rm for}\;N\;{\rm odd}  \\
(- i)^{\frac{N-2}{2}}  & {\rm for}\;N \;{\rm even} \end{array} \right.\! .
\end{equation}

\ni Thus, in the case od $N$ odd we can represent the Dirac-type current 
(20) in the form

%26
\begin{equation}
j^{(N)}_{\mu {\rm D}}(x) = \psi^{(N) *}_{\alpha_1 \alpha_2 \ldots 
\alpha_N}(x)  (\gamma_0  \gamma_\mu)_{\alpha_1\, \beta_1} 
(\gamma_0)_{\alpha_2\, \beta_2} \ldots (\gamma_0)_{\alpha_N\, \beta_N} 
\psi^{(N)}_{\beta_1 \beta_2 \ldots \beta_N}(x) \,.
\end{equation}

\ni From Eqs. (18) and (26) the relationship (21) follows.

For $N$ odd, the Dirac-type current (20) or (26) is locally conserved, but 
such is also the relativistic noncovariant structure $j^{(N)}_{\mu 
\alpha_2 \ldots \alpha_N , \alpha_2 \ldots \alpha_N}(x)$ calculated from 
the definition (18) by summing over $ \alpha_2 = \beta_2,\ldots,\alpha_N = 
\beta_N$. It follows that the Hermitian, not explicitly covariant 
$N\times N$ bispinor matrix

%27
\begin{equation}
\xi^{(N)} \Gamma^{(N)}_{20} \ldots \Gamma^{(N)}_{N0} = {\bf 1}\otimes 
\,\underbrace{\gamma_0 \, \otimes \gamma_0  \otimes \ldots \otimes 
\gamma_0 }_{N-1\;{\rm times}} = \left( \delta_{\alpha_1 \beta_1} 
(\gamma_0)_{\alpha_2 \beta_2}  \ldots (\gamma_0)_{\alpha_N \beta_N} \right) 
\end{equation}

\ni is a constant of motion. This matrix may be called the total 
"relative"\, internal parity. Imposing on $\psi^{(N)}(x) \equiv 
\left( \psi^{(N)}_{\alpha_1 \alpha_2 \ldots \alpha_N}(x) \right) $ 
the stationary constraint in the form of the eigenvalue equation

%28
\begin{equation}
\xi^{(N)} \Gamma^{(N)}_{20} \ldots \Gamma^{(N)}_{N0} \psi^{(N)}(x) = 
\psi^{(N)}(x) \,,
\end{equation}

\ni requiring that the eigenvalue of total "relative"\, internal parity 
must be always equal to +1, we simplify the relativistic covariant 
Dirac-type current (20) or (26) to the form

%29
\begin{eqnarray}
j^{(N)}_{\mu {\rm D}} & = & \psi^{(N) *}_{\alpha_1 \alpha_2 \ldots 
\alpha_N}(x) \left( \Gamma^{(N)}_{10} \Gamma^{(N)}_{1\mu} 
\right)_{\alpha_1 \alpha_2 \ldots \alpha_N, \beta_1 \beta_2 \ldots 
\beta_N} \psi^{(N)}_{\beta_1 \beta_2 \ldots \beta_N}(x) \nonumber \\ 
& = & \psi^{(N) *}_{\alpha_1 \alpha_2 \ldots \alpha_N}(x) 
\left( \gamma_0 \gamma_\mu \right)_{\alpha_1 \beta_1} 
\psi^{(N)}_{\beta_1 \alpha_2 \ldots \alpha_N}(x)
\end{eqnarray}

\ni that is not explicitly covariant in the world of "relative"\, Dirac 
degrees of freedom. The form (29) leads to the positive-definiteness 
of $\psi^{(N)}(x)$,

%30
\begin{equation}
j^{(N)}_{0 {\rm D}}(x) = \psi^{(N) *}_{\alpha_1 \alpha_2 \ldots 
\alpha_N}(x) \psi^{(N)}_{\alpha_1 \alpha_2 \ldots \alpha_N}(x) > 0\,,
\end{equation}

\ni which is a natural requirement for $\psi^{(N)}(x)$.

It is not difficult to demonstrate that wave functions or fields 
of spin-1/2 fundamental fermions, $\psi^{(f_N)}_{\alpha_1}(x) \, 
(N = 1,3,5)$ defined in Eqs. (14), satisfy the constraint (28). 
In fact, writing  

%31
\begin{equation}
\psi^{(3)}_{\alpha_1 \alpha_2 \alpha_3}(x) = (\gamma_5 C)_{\alpha_3 
\alpha_2} \psi^{(f_3)}_{\alpha_1}(x) \;\;,\;\;\psi^{(5)}_{\alpha_1 
\alpha_2 \alpha_3 \alpha_4 \alpha_5}(x) = \varepsilon_{\alpha_2 
\alpha_3 \alpha_4 \alpha_5} \psi^{(f_5)}_{\alpha_1}(x)\,, 
\end{equation}

\ni we check that

%32
\begin{equation}
\delta_{\alpha_1 \beta_1} (\gamma_0)_{\alpha_2 \beta_2} 
(\gamma_0)_{\alpha_3 \beta_3}  \psi^{(3)}_{\beta_1 \beta_2 \beta_3}(x)
= \psi^{(3)}_{\alpha_1 \alpha_2 \alpha_3}(x)
\end{equation}

\ni and

%33
\begin{equation}
\delta_{\alpha_1 \beta_1} (\gamma_0)_{\alpha_2 \beta_2} 
(\gamma_0)_{\alpha_3 \beta_3} (\gamma_0)_{\alpha_4 \beta_4} 
(\gamma_0)_{\alpha_5 \beta_5} \psi^{(5)}_{\beta_1 \beta_2 \beta_3 
\beta_4 \beta_5}(x) = \psi^{(5)}_{\alpha_1 \alpha_2 \alpha_3 
\alpha_4 \alpha_5}(x)\,, 
\end{equation}

\ni where in the chiral representation

%34
\begin{equation}
\gamma_5 = \left( \begin{array}{rrrr} 1 & 0 & 0 & 0 \\ 0 & 1 & 0 & 0 \\ 
0 & 0 & -1 & 0  \\ 0 & 0 & 0 & -1 \end{array}\right)\;,\; C = 
\left( \begin{array}{rrrr} 0 & -1 & 0 & 0 \\ 1 & 0 & 0 & 0 \\ 
0 & 0 & 0 & 1   \\  0 & 0 & -1 & 0 \end{array}\right) \;,\;\gamma_0 = 
\left( \begin{array}{rrrr} 0 & 0 & 1 & 0 \\ 0 & 0 & 0 & 1 \\ 1 & 0 & 0 & 0  
\\ 0 & 1 &  0 & 0  \end{array}\right)\;.
\end{equation}%w 268?

\ni Here, $(\gamma_5 C)^T = -\gamma_5 C$.

In conclusion, we would like to emphasize that the phenomenon of existence 
in Nature of three generations of fundamental fermions (leptons and quarks) 
can be understood in a satisfactory way on the base of two postulates:

\vspace{0.3cm}

\ni ({\it i}) For all fundamental particles of matter the Dirac square-root 
procedure $ \sqrt{p^2} \rightarrow \Gamma^{(N)} \cdot p$ works, leading in 
the interaction-free case to the sequence $N = 1,2,3,\ldots$ of Dirac-type 
equations (4), satisfied by the sequence $N = 1,2,3,\ldots$ of wave 
functions or fields $\psi^{(N)}(x) \equiv \left(\psi^{(N)}_{\alpha_1 
\alpha_2 \ldots \alpha_N}(x)  \right)$, where $\alpha_1$ is a 
"centre-of-mass"\, Dirac bispinor index and $\alpha_2, \ldots, 
\alpha_N $ are "relative"\, Dirac bispinor indices.

\vspace{0.3cm}

\ni ({\it ii}) The "centre-of-mass"\, Dirac bispinor index $\alpha_1$ 
is coupled to the \SM gauge fields through the term $ g \gamma_{\alpha_1 
\beta_1} \cdot A(x) $ in the Dirac-type equation (13), while the 
"relative"\, Dirac bispinor indices $\alpha_2, \ldots, \alpha_N $ are all 
free undistinguishable physical objects obeying Fermi statistics along with 
the Pauli principle (called then "intrinsic Pauli principle") which requires 
the full antisymmetry of $\psi^{(N)}_{\alpha_1 \alpha_2 \ldots \alpha_N}(x) $
 with respect to $\alpha_2, \ldots, \alpha_N$.

\vspace{0.3cm}

\ni This antisymmetry allows only for {\it three} $N = 1,3,5$  in the
 case of $N$ odd, and for {\it two} $N = 2,4$  in the case of $N$ even. 
Hence, unavoidably, there follow {\it three and only three} generations 
of fundamental fermions (leptons and quarks) and {\it two and only two} 
generations of some fundamental bosons (not recognized yet). The former 
carry spin 1/2 and the latter spin 0 (as was argued in Ref. \cite{KR}). All 
possess a conventional \SM signature, suppressed in our notation.

The Dirac-type matrices $ \Gamma^{(N)}_\mu$, appearing in the square-root 
procedure $ \sqrt{p^2} \rightarrow \Gamma^{(N)} \cdot p$, are here 
constructed by means of the Clifford algebra (3) of matrices 
$\gamma^{(N)}_{i \mu}$, and then embedded into of the Clifford algebra (5) 
of matrices $ \Gamma^{(N)}_{i \mu}$ {\it via} a Jacobi definition (6), where 
$\Gamma^{(N)}_{1 \mu} \equiv \Gamma^{(N)}_\mu$ and 
$\Gamma^{(N)}_{2 \mu},\ldots,\Gamma^{(N)}_{N \mu}$ play the role of  
"centre-of-mass"\, and "relative"\, Dirac-type matrices, respectively, 
defining one "centre-of-mass"\, and $N-1$ "relative" Dirac bispinor 
indices ($\alpha_1$ and $\alpha_2, \ldots,\alpha_N$).

\vfill\eject

\baselineskip 0.72cm

\vspace{0.6cm}

{
%\vfill\eject


\centerline{\bf References}}

\vspace{0.25cm}


\begin{thebibliography}{99}

\bibitem{DIR} {\it Cf.} P.A.M. Dirac, {\it Principles of Quantum Mechanics}, 
              4$^{th}$ edition, Oxford University Press 1958, \S 67.

\bibitem{KR} W. Kr\'{o}likowski, {\it Acta Phys. Pol.} {\bf B21}, 871 (1990); 
             {\it Phys. Rev.} {\bf D45}, 3222 (1992); {\bf D46}, 5188 (1992);
             in {\it Spinors, Twistors, Clifford Algebras and Quantum 
             Deformations (Proc. 2nd Max Born Symposium 1992)}, eds. 
             Z.~Oziewicz {\it et al.}, Kluwer Acad. Press 1993; 
             {\it Acta Phys. Pol.} {\bf B24}, 1149 (1993); {\bf B27}, 2121 
             (1996); {\it cf.} also Appendices in hep-ph/0108157 and 
             hep-ph/0201004v2.

\bibitem{BA} T. Banks, Y. Dothan and D.~Horn, {\it Phys. Lett.} {\bf B 117}, 
             413 (1982).

\bibitem{KA} E. K\"{a}hler, {\it Rendiconti di Matematica} {\bf 21}, 425 
             (1962); {\it cf.} also D.~Ivanenko and L.~Landau, {\it Z. Phys.}
             {\bf 48}, 341 (1928).

\end{thebibliography}
\end{document}